\documentclass[runningheads]{llncs}
\usepackage{graphicx}

\usepackage{booktabs} 
\usepackage{color}
\usepackage{multirow}
\usepackage{subfigure}
\usepackage[algo2e]{algorithm2e}
\usepackage{algorithmic}
\usepackage{algorithm}
\usepackage{epstopdf}

\usepackage{amsmath,amssymb,amsfonts}
\usepackage{textcomp}
\usepackage{xcolor}
\usepackage{amsmath}

\usepackage{fix-cm}
\def\BibTeX{{\rm B\kern-.05em{\sc i\kern-.025em b}\kern-.08em
    T\kern-.1667em\lower.7ex\hbox{E}\kern-.125emX}}

\begin{document}

\title{ Data Centers Job Scheduling with Deep Reinforcement Learning}

\author{Sisheng Liang \and
Zhou Yang \and
Fang Jin \and
Yong Chen \\}

\institute{
Department of Computer Science\\
Texas Tech University\\
\email{\{sisheng.liang, zhou.yang, fang.jin, yong.chen\}@ttu.edu}\\}


\maketitle
\begin{abstract}
Efficient job scheduling on data centers under heterogeneous complexity is crucial but challenging since it involves the allocation of multi-dimensional resources over time and space. 
To adapt the complex computing environment in data centers, we proposed an innovative Advantage Actor-Critic (A2C) deep reinforcement learning based approach called A2cScheduler for job scheduling. A2cScheduler consists of two agents, one of which, dubbed the actor, is responsible for learning the scheduling policy automatically and the other one, the critic, reduces the estimation error. Unlike previous policy gradient approaches, A2cScheduler is designed to reduce the gradient estimation variance and to update parameters efficiently.
We show that the A2cScheduler can achieve competitive scheduling performance using both simulated workloads and real data collected from an academic data center.


\end{abstract}

\section{Introduction}

Job scheduling is a critical and challenging task for computer systems since it involves a complex allocation of limited resources such as CPU/GPU, memory and IO among numerous jobs. It is one of the major tasks of the scheduler in a computer system's Resource Management System (RMS), especially in high-performance computing (HPC) and cloud computing systems, where inefficient job scheduling may result in a significant waste of valuable computing resources. 
Data centers, including HPC systems and cloud computing systems, have become progressively more complex in their architecture~\cite{van2012scheduling}, configuration(e.g., special visualization nodes in a cluster)~\cite{hovestadt2003scheduling} and the size of work and workloads received~\cite{garg2011sla}, all of which increase the job scheduling complexities sharply.

The undoubted importance of job scheduling has fueled interest in the scheduling algorithms on data centers. At present, the fundamental scheduling methodologies~\cite{zhou2013exploring}, such as FCFS (first-come-first-serve), backfilling, and priority queues that are commonly deployed in data centers are extremely hard and time-consuming to configure, severely compromising system performance, flexibility and usability. To address this problem, several researchers have proposed data-driven machine learning methods that are capable of automatically learning the scheduling policies, thus reducing human interference to a minimum. Specifically, a series of policy based deep reinforcement learning approaches have been proposed to manage CPU and memory for incoming jobs~\cite{Mao:2016:RMD:3005745.3005750}, schedule time-critical workloads~\cite{liu2018reinforcement}, handle jobs with dependency~\cite{mao2018learning}, and schedule data centers with hundreds of nodes~\cite{domeniconi2019cush}. 
Despite the extensive research into job scheduling, however, the increasing heterogeneity of the data being handled remains a challenge. These difficulties arise from multiple issues. First, policy gradient DRL method based scheduling method suffers from a high variance problem, which can lead to low accuracy when computing the gradient.  
Second, previous work has relied on used Monte Carlo (MC) method to update the parameters, which involved massive calculations, especially when there are large numbers of jobs in the trajectory.

To solve the above-mentioned challenges, we propose a policy-value based deep reinforcement learning scheduling method called A2cScheduler, which can satisfy the heterogeneous requirements from diverse users, improve the space exploration efficiency, and reduce the variance of the policy.
A2cScheduler consists of two agents named actor and critic respectively, the actor is responsible for learning the scheduling policy and the critic reduces the estimation error. 
The approximate value function of the critic is incorporated as a baseline to reduce the variance of the actor, thus reducing the estimation variance considerably~\cite{sutton1998introduction}. A2cScheduler updates parameters via the multi-step Temporal-difference (TD) method, which speeds up the training process markedly compared to conventional MC method due to the way TD method updates parameters. 
The main contributions are summarized as below:
\begin{enumerate}
      \item This represents the first time that A2C deep reinforcement has been successfully applied to a data center resource management, to the best of the authors' knowledge. 
    \item A2cScheduler updates parameters via multi-step Temporal-difference (TD) method which speeds up the training process comparing to MC method due to the way TD method updates parameters. This is critical for the real world data center scheduling application since jobs arrive in real time and low latency is undeniably important.       
    \item We tested the proposed approach on both real-world and simulated datasets, and results demonstrate that our proposed model outperformed many existing widely used methods.  
\end{enumerate}

\section{Related Work}

\paragraph{\textbf{Job scheduling with deep reinforcement learning}}
Recently, researchers have tried to apply deep reinforcement learning on cluster resources management. A resource manager DeepRM was proposed in~\cite{Mao:2016:RMD:3005745.3005750} to manage CPU and memory for incoming jobs. The results show that policy based deep reinforcement learning outperforms the conventional job scheduling algorithms such as Short Job First and Tetris~\cite{grandl2015multi}.
\cite{liu2018reinforcement} improves the exploration efficiency by adding baseline guided actions for time-critical workload job scheduling.
\cite{yangcoordinating} discussed heuristic based method to coordinate disaster response. 
Mao proposed Decima in~\cite{mao2018learning} which could handle jobs with dependency when graph embedding technique is utilized. \cite{domeniconi2019cush} proved that policy gradient based deep reinforcement learning can be implemented to schedule data centers with hundreds of nodes. 

\paragraph{\textbf{Actor-critic reinforcement learning}}
Actor-critic algorithm is the most popular algorithm applied in the reinforcement learning framework~\cite{grondman2012survey} which falls into three categories: actor-only, critic-only and actor-critic methods~\cite{konda2000actor}. Actor-critic methods combine the advantages of actor-only
and critic-only methods. Actor-critic methods usually have good convergence properties, in contrast to critic-only~\cite{grondman2012survey}. 
At the core of several recent state-of-the-art Deep RL algorithms is the advantage actor-critic (A2C) algorithm~\cite{mnih2016asynchronous}. In addition to learning a policy (actor) $\pi(a | s ; \theta)$, A2C learns a parameterized critic: an estimate of value function $v_{\pi}(s)$, which then uses both to estimate the remaining return after k steps, and as a control variate (i.e. baseline) that reduces the variance of the return estimates~\cite{srinivasan2018actor}.
\section{Method and Problem Formulation}
In this section, we first review the framework of A2C deep reinforcement learning, and then explain how the proposed A2C based A2cScheduler works in the job scheduling on data centers. The rest part of this section covers the essential details about model training. 

\begin{figure*}[!t]
 \centering
 \includegraphics[width=0.8\linewidth]{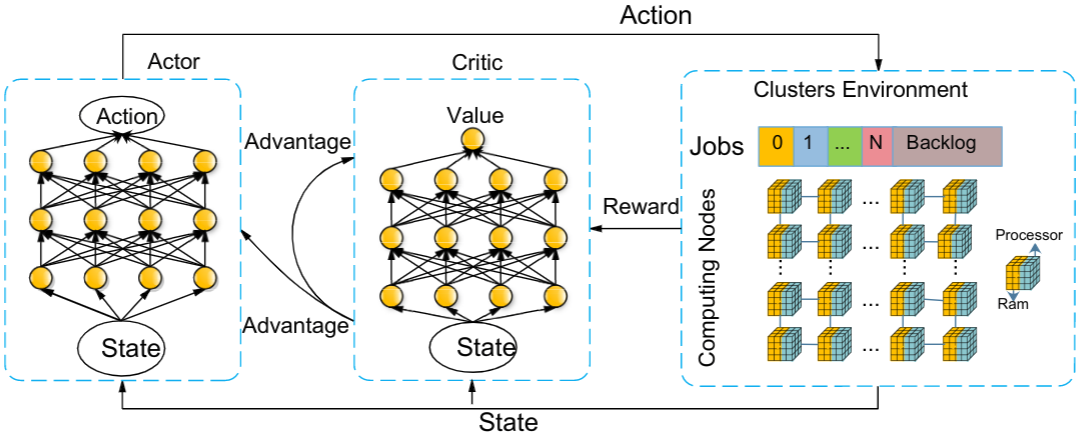}
 \caption{A2cScheduler job scheduling framework.}
 \label{fig:framework} 
\end{figure*}

\subsection{A2C in Job Scheduling}

The Advantage Actor-critic (A2C), which combines policy based method and value based method, can overcome the high variance problem from pure policy gradient approach.
The A2C algorithm is composed of a policy $\pi\left(a_{t} | s_{t} ; \theta\right)$ and a value function $V\left(s_{t} ; w\right)$, where policy is generated by policy network and value is estimated by critic network. 
The proposed the A2cScheduler framework is shown in figure~\ref{fig:framework}, which consists of an actor network, a critic network and the cluster environment. The cluster environment includes a global queue, a backlog and the simulated machines. The queue is the place holding the waiting jobs. The backlog is an extension of the queue when there is not enough space for waiting jobs. Only jobs in the queue will be allocated in each state.

\paragraph{\textbf{The setting of A2C}}
\begin{itemize}
    \item \textbf{Actor}: The policy $\pi$ is an actor which generates probability for each possible action. $\pi$ is a mapping from state $s_t$ to action $a_t$. Actor can choose a job from the queue based on the action probability generated by the policy $\pi$. For instance, given the action probability $P=\{p_1,\dots, p_{N}\}$ for N actions, $p_i$ denotes the probability that action ${a}_i$ will be selected.
    If the action is chosen according to the maximum probability ($action=$ $\operatorname*{arg\,max}_{i \in [0,N], i \in N^+}{p_i}$), the actor acts greedily which limits the exploration of the agent. Exploration is allowed in this research. The policy is estimated by a neural network $\pi(a|s,\theta)$, where $a$ is an action, $s$ is the state of the system and $\theta$ is the weights of the policy network.
    \item \textbf{Critic}: A state-value function $v(s)$ used to evaluate the performance of the actor. It is estimated by a neural network $\hat{v}(s, \mathbf{w})$ in this research where $s$ is the state and $\mathbf{w}$ is the weights of the value neural network. 
    \item \textbf{State $s_t \in S$}: A state $s_t$ is defined as the resources allocation status of the data center including the status of the cluster and the status of the queue at time $t$. The states $S$ is a finite set. Figure~\ref{fig:scheduling3} shows an example of the state in one time step. The state includes three parts: status of the resources allocated and the available resources in the cluster, resources requested by jobs in the queue, and status of the jobs waiting in the backlog. The scheduler will only schedules jobs in the queue. 
    \item \textbf{Action $a_t \in A$}: An action $a_t=\{a_t\}_1^{N}$ denotes the allocation strategy of jobs waiting in the queue at time $t$, where $N$ is the number of slots for waiting jobs in the queue. The action space $A$ of an actor specifies all the possible allocations of jobs in the queue for the next iteration, which gives a set of $N+1$ discrete actions represented by $\{\emptyset, 1, 2,\dots,N \}$ where $a_t =i$( $\forall i \in \{1,\dots, N\}$) means the allocation of the $i^{th}$ job in the queue and $a_t=\emptyset$ denotes a void action where no job is allocated. 
  
  \begin{figure}[th]
 \centering
 \includegraphics[width=0.55\linewidth]{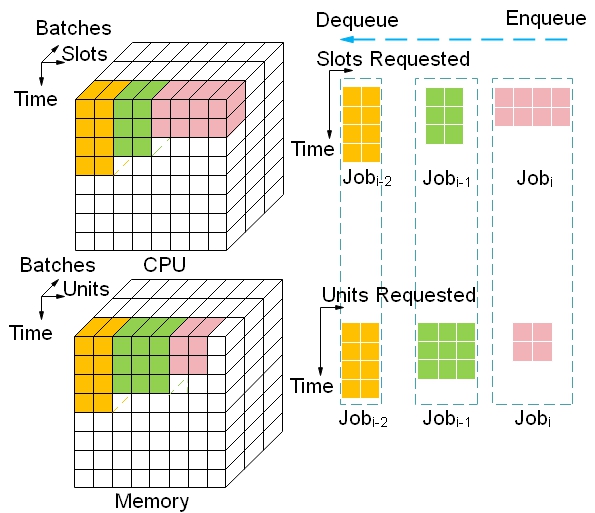}
 \caption{An example of the tensor representation of a state. At each iteration, the decision combination of number of jobs will be scheduled is $2^{Total_{jobs}}$, which has exponential growth rate. We simplify the case by selecting a decision from $decision\_domain=\{0, 1, \dots, N\}$, where $N$ is a fixed hyper-parameter, $decision=i$ denotes select $i^{th}$ job, and $decision=0$ denotes no job will be selected.}
 \label{fig:scheduling3} 
\end{figure}

     \item \textbf{Environment}: The simulated data center contains resources such as CPUs, RAM and I/O. It also includes resource management queue system in which jobs are waiting to be allocated. 
     
    
    \item \textbf{Discount Factor $\gamma$}: A discount factor $\gamma$ is between 0 and 1, and is used to quantify the difference in importance between immediate rewards and future rewards. The smaller of $\gamma$, the less importance of future rewards. 
    
    \item \textbf{Transition function $P: S \times A \to [0,1]$}: Transition function describes the probabilities of moving between current state to the next state. The state transition probability $p(s_{t+1}|s_t, a_t)$ represents the probability of transiting to $s_{t +1} \in S$ given a joint action $a_t \in A$ is taken in the current state $s_t \in S$. 
    
    \item \textbf{Reward function $r \in R = S \times A \to (-\infty, +\infty)$}: A reward in the data center scheduling problem is defined as the feedback from the environment when the actor takes an action at a state. The actor attempts to maximize its expected discounted reward:

    ${R_t}=E({r_t^i}+{\gamma}{r_{t+1}^i}+...)=E({\sum\limits_{k=0}^{\infty}{\gamma}^k}{r_{t+k}^i})=E({r_t^i}+{\gamma}{R_{t+1}})$. 
    
    The agent reward at time $t$ is defined as 
    $r_t=-\frac{1}{T_j}$ 
    , where $T_j$ is the runtime for job $j$. 

\end{itemize}

The goal of data center job scheduling is to find the optimal policy $\pi^*$ (a sequence of actions for agents) that maximizes the total reward. The state value function ${Q^\pi}(s,a)$ is introduced to evaluate the performance of different policies. ${Q^\pi}(s,a)$ stands for the expected total reward with discount from current state $s$ on-wards with the policy $\pi$, which is equal to:
\begin{equation}
\begin{split}
{Q^{\pi}}(s_t,a_t) = &{E_{\pi}}({R_t}|{s_t, a_t}) = {E_{\pi}}({r_t}+{\gamma}{Q^{\pi}}({s',a'}))\\
 =& {r_t}+\gamma{{\sum\limits_{{s'} \in S}}{P^{\pi}}({s'}|s){Q^{\pi}}({s',a'})}
\end{split}
\end{equation}

, where $s'$ is the next state, and $a'$ is the action for the next time step. 

Function approximation is a way for generalization when the state and/or action spaces are large or
continuous. Several reinforcement learning algorithms have been proposed to estimate the value of an action in various contexts such as the Q-learning~\cite{watkins1992q} and SARSA~\cite{sprague2003multiple}. Among them, the model-free Q-learning algorithm stands out for its simplicity~\cite{al2007model}. In Q-learning, the algorithm uses a Q-function to calculate the total reward, defined as $Q: S\times A \to R$. Q-learning iteratively evaluates the optimal Q-value function using backups:
\begin{equation}
Q(s,a)=Q(s,a)+\alpha[r+\gamma{max_{a'}}Q(s',a')-Q(s,a)]
\end{equation}, where $\alpha \in [0,1)$ is the learning rate and the term in the brackets is the temporal-difference (TD) error. Convergence to $Q^{\pi^*}$ is guaranteed in the tabular case provided there is sufficient state/action space exploration. 
\paragraph{\textbf{The loss function for critic}} Loss function of the critic is utilized to update the critic network parameters.
\begin{equation}\label{eq:critic_loss}
L(w_i)=\mathbb{E}(r+\gamma{max_{a'}}Q(s',a';w_{i-1})-Q(s,a;w_{i}))^2,
\end{equation}
where $s'$ is the state encountered after state $s$. Critic update the parameters of the value network by minimizing critic loss in equation \ref{eq:critic_loss}. 

\paragraph{\textbf{Advantage actor-critic}}
The critic updates state-action value function parameters, and the actor updates policy parameters, in the direction suggested by the critic. A2C updates both the policy and value-function networks with the multi-step returns as described in \cite{mnih2016asynchronous}. Critic is updated by minimizing the loss function of equation \ref{eq:critic_loss}. Actor network is updated by minimizing the actor loss function in equation 
\begin{equation}\label{eq:actor_loss}
L(\theta_i^{\prime})=\nabla_{\theta^{\prime}} \log \pi\left(a_{t} | s_{t} ; \theta^{\prime}\right) A\left(s_{t}, a_{t} ; \theta, w_i\right)
\end{equation}, where $\theta_i$ is the parameters of the actor neural network and $w_i$ is the parameters of the critic neural network. Note that the parameters $\theta_i$ of policy and $w_i$ of value are distinct for generality. Algorithm \ref{A2C} presents the calculation and update of parameters per episode.

\subsection{Training algorithm}
The A2C consists of an actor and a critic, and we implement both of them using deep
convolutional neural network. For the Actor neural network, it takes the afore-mentioned tensor
representation of resource requests and machine status as the input, and outputs the probability distribution over
all possible actions, representing the jobs to be scheduled. For the Critic neural network, it
takes as input the combination of action and the state of the system, and
outputs the a single value, indicating the evaluation for actor's action. 


\begin{algorithm}[htbp]

    \KwIn{a policy parameterization $\pi(a|s,\theta)$}
    \KwIn{ a state-value function parameterization $\hat{v}(s, \mathbf{w})$}
    Parameters: step sizes {  $\alpha^{\theta}>0, \alpha^{\mathrm{w}}>0$}\
    
    Initialization: policy parameter $\boldsymbol{\theta} \in \mathbb{R}^{d^{\prime}}$ and state-value function weights   $\mathbf{w} \in \mathbb{R}^{d}(\text { e.g. }, \text { to } \mathbf{0.001})$
    
    \KwOut{The scheduled sequence of jobs[1..n]}
    
    Loop forever (for each episode):
    
    \Indp
    Initialize S (state of episode)    
    
    Loop while S is not terminal~(for each time step of episode):
    
    \Indp
    $A \sim \pi(\cdot | S, \boldsymbol{\theta})$
    
    Take action $A$, observe state S$^{\prime}, reward~R$
    
    $\delta \leftarrow R+\gamma \hat{v}\left(S^{\prime},\mathbf{w}\right)-\hat{v}(S, \mathbf{w})  ~~(\text { If }~S^{\prime} \text { is terminal, then } \hat{v}\left(S^{\prime}, \mathbf{w}\right) \doteq 0)$
    
    $\mathbf{w} \leftarrow \mathbf{w}+\alpha^{\mathbf{w}} \delta \nabla \hat{v}(S, \mathbf{w})$
    
    $\boldsymbol{\theta} \leftarrow \boldsymbol{\theta}+\alpha^{\boldsymbol{\theta}} \delta \nabla \ln \pi(A | S, \boldsymbol{\theta})$
    
    $S \leftarrow S^{\prime}$
    \caption{A2C reinforcement learning scheduling algorithm }
    \label{A2C}
\end{algorithm}

\begin{table*}[!thbp]
\tiny
  \centering
  \caption{{Performance comparison when model converged.}}
    \begin{tabular}{p{6.7em}p{5.1em}p{5.1em}p{5.1em}p{5.1em}p{5.1em}p{5.1em}p{5.1em}p{5.1em}}
    \toprule\toprule
    \multicolumn{9}{c}{Job Rate} \\
    \cline{2-5} \cline{6-9}
    \multicolumn{1}{l}{} & \multicolumn{4}{c}{0.9}       & \multicolumn{4}{c}{0.8} \\
    \midrule
    Type  & \multicolumn{1}{l}{Random} & \multicolumn{1}{l}{Tetris} & \multicolumn{1}{l}{SJF} & \multicolumn{1}{l}{A2cScheduler} & \multicolumn{1}{l}{Random} & \multicolumn{1}{l}{Tetris} & \multicolumn{1}{l}{SJF} & \multicolumn{1}{l}{A2cScheduler} \\
    \midrule
    Slowdown & 5.50$\pm$0.00 & 2.90$\pm$0.00 & \textbf{1.81$\pm$0.00} & 2.03$\pm$0.01 & 6.2$\pm$0.00 & 3.25$\pm$0.00 & 2.52$\pm$0.00 & \textbf{2.30$\pm$0.05} \\
    \midrule
    Complete time & 12.51$\pm$0.00 & 8.61$\pm$0.00 & 7.42$\pm$0.00 & \textbf{7.20$\pm$0.01} & 14.21$\pm$0.00 & 8.50$\pm$0.00 & 6.50$\pm$0.00 & \textbf{6.20$\pm$0.04} \\
    \midrule
    Waiting time & 8.22$\pm$0.00 & 3.32$\pm$0.00 & 2.21$\pm$0.00 & \textbf{2.20$\pm$0.01} & 9.15$\pm$0.00 & 2.10$\pm$0.00 & \textbf{1.93$\pm$0.00} & 2.12$\pm$0.005 \\
    \bottomrule\bottomrule
    \end{tabular}%
  \label{tab:converged_98}%
\end{table*}%


\section{Experiments}
\subsection{Experiment Setup} 

The experiments are executed on a desktop computer with two RTX-2080 GPUs and one i7-9700k 8-core CPU. A2cScheduler is implemented using Tensorflow framework. 
Simulated jobs arrive online in Bernouli process. A piece of job trace from a real data center is also tested. CPU and Memory are the two kinds of resources considered in this research. 

The training process begins with an initial state of the data center. At each time step, a state is passed into the policy network $\pi$. An action is generated under policy $\pi$. A void action is made or a job is chosen from the global queue and put into the cluster for execution. Then a new state is generated and some reward is collected. The states, actions, policy and rewards are collected as trajectories. Meanwhile, the state is also passed into the value network to estimate the value, which used to evaluate the performance of the action. Actor in A2cScheduler learns to produce resource allocation strategies from experiences after epochs.

\begin{table*}[!thbp]
\tiny
  \centering
  \caption{{Performance comparison when model converged.}}
    \begin{tabular}{p{6.95em}p{5.1em}p{5.1em}p{5.1em}p{5.1em}p{5.1em}p{5.1em}p{5.1em}p{5.1em}}
    \toprule\toprule
    \multicolumn{9}{c}{Job Rate} \\
    \cline{2-5} \cline{6-9}
    \multicolumn{1}{l}{} & \multicolumn{4}{c}{0.7}       & \multicolumn{4}{c}{0.6} \\
    \midrule
    Type  & \multicolumn{1}{l}{Random} & \multicolumn{1}{l}{Tetris} & \multicolumn{1}{l}{SJF} & \multicolumn{1}{l}{A2cScheduler} & \multicolumn{1}{l}{Random} & \multicolumn{1}{l}{Tetris} & \multicolumn{1}{l}{SJF} & \multicolumn{1}{l}{A2cScheduler} \\
    \midrule
    Slowdown & 5.05$\pm$0.00 & 3.32$\pm$0.00 & 2.14$\pm$0.00 & \textbf{1.91$\pm$0.02} & 3.22$\pm$0.00 & 1.82$\pm$0.00 & 1.56$\pm$0.00 & \textbf{1.36$\pm$0.04} \\
    \midrule
    Complete time & 13.15$\pm$0.00 & 10.02$\pm$0.00 & 7.66$\pm$0.00 & \textbf{6.10$\pm$0.03} & 10.0$\pm$0.00 & 5.50$\pm$0.00 & 5.50$\pm$0.00 & 5.50$\pm$0.04 \\
    \midrule
    Waiting time & 8.32$\pm$0.00 & 4.51$\pm$0.00 & 2.53$\pm$0.00 & \textbf{1.82$\pm$0.03} & 8.32$\pm$0.00 & 1.48$\pm$0.00 & 1.48$\pm$0.00 & 1.50$\pm$0.003 \\
    \bottomrule\bottomrule
    \end{tabular}%
  \label{tab:converged_76}%
\end{table*}%

\subsection{Evaluation Metrics} 
Reinforcement learning algorithms, including A2C, have been mostly evaluated by converging speed. However, these metrics are not very informative in domain-specific applications such as scheduling. Therefore, we present several evaluation metrics that are helpful for access the performance of the proposed model.

Given a set of jobs $J=\{j_1, \dots, j_N\}$, where $i^{th}$ job is associated with arrival time $t_i^{a}$, finish time $t_i^{f}$, and execution time $t_i^{e}$.
\paragraph{Average job slowdown}  The slowdown for $i^{th}$ job is defined as $s_i = \frac{t_i^{f}-t_i^{a}}{t_i^{e}}=\frac{c_i}{t_i}$, where $c_i=t_i^{f}-t_i^{a}$ is the completion time of the job and $t_i$ is the duration of the job. The average job slowdown is defined as $s_{avg} =\frac{1}{N}\sum\limits_{i=1}^n \frac{t_i^{f}-t_i^{a}}{t_i^{e}}= \frac{1}{n}\sum\limits_{i=1}^N \frac{c_i}{t_i}$. The slowdown metric is important because it helps to evaluate normalized waiting time of a system.

\paragraph{Average job waiting time} For the $i^{th}$ job, the waiting time $t_{wi}$ is the time between arrival and start of execution, which is formally defined as $t_{wi}=t_i^{s}-t_i^{a}$.


    



\begin{figure*}[t]
    \centering
	\subfigure[Discounted reward.]{
	    \includegraphics[width=0.4\linewidth]{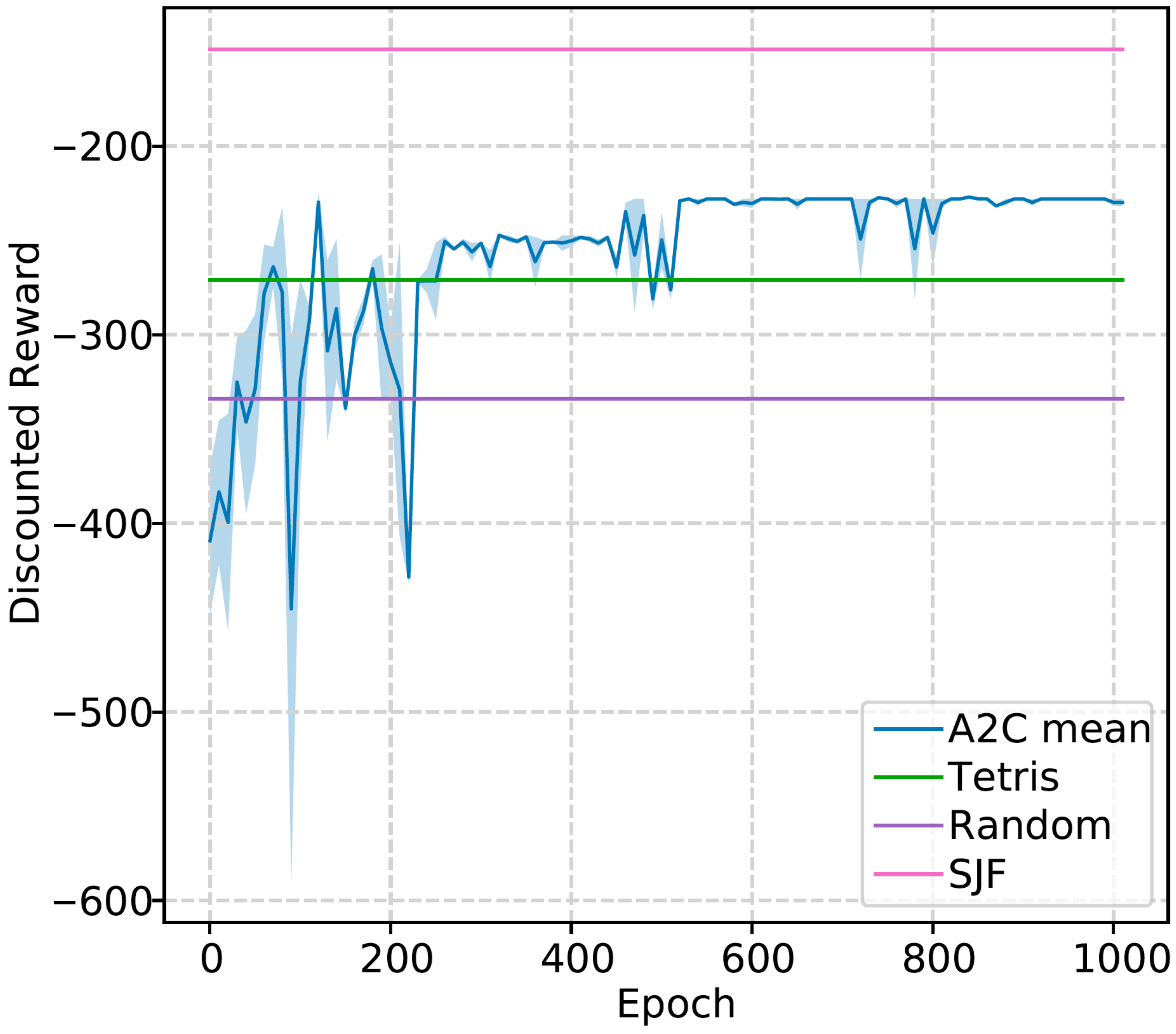}
	    \label{fig:learning_reward_0.9}
	}
	\subfigure[Slowdown.]{
	    \includegraphics[width=0.4\linewidth]{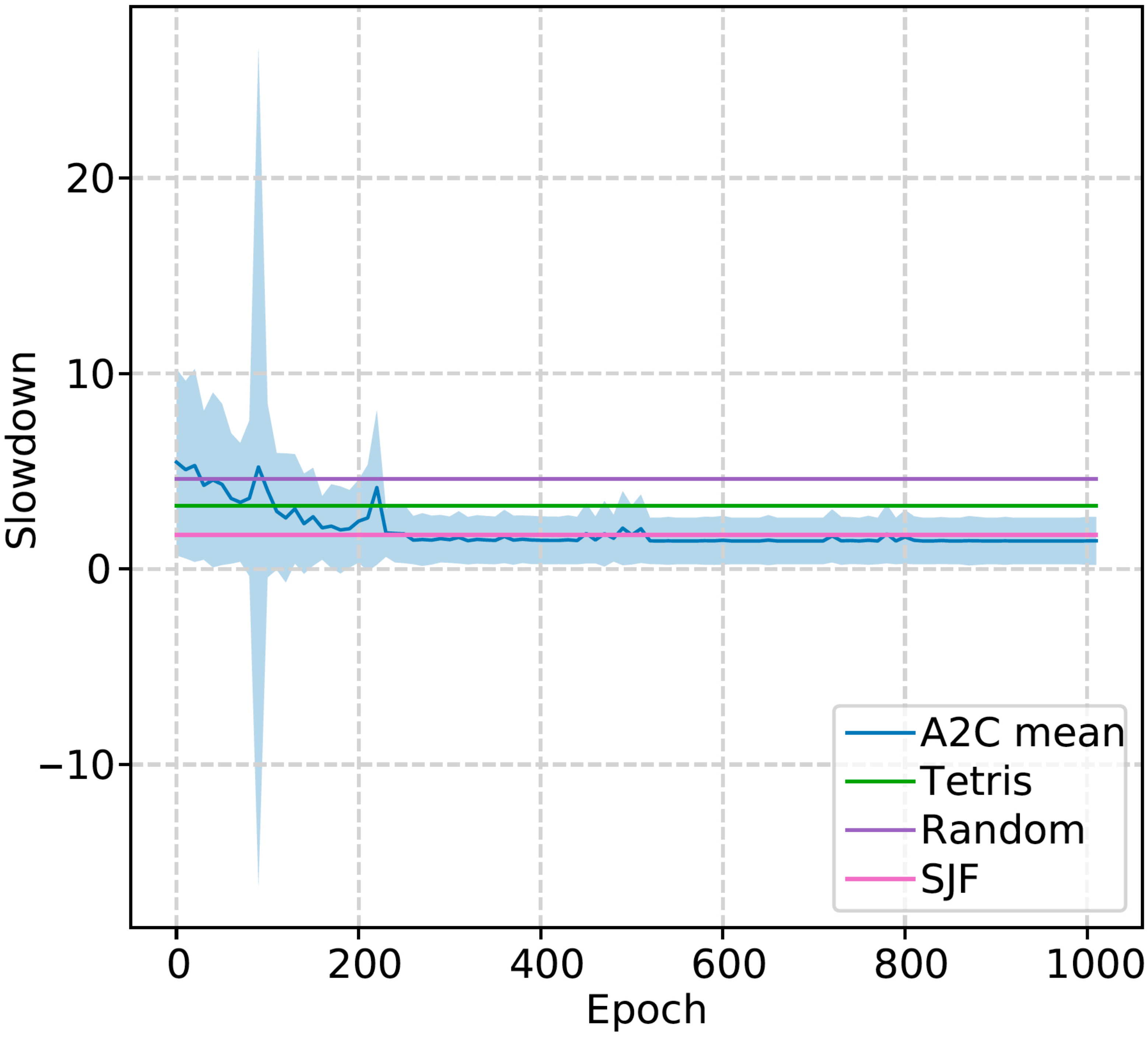}
	    \label{fig:learning_slown_down0.9}
	}
	\subfigure[Average completion time.]{
	    \includegraphics[width=0.4\linewidth]{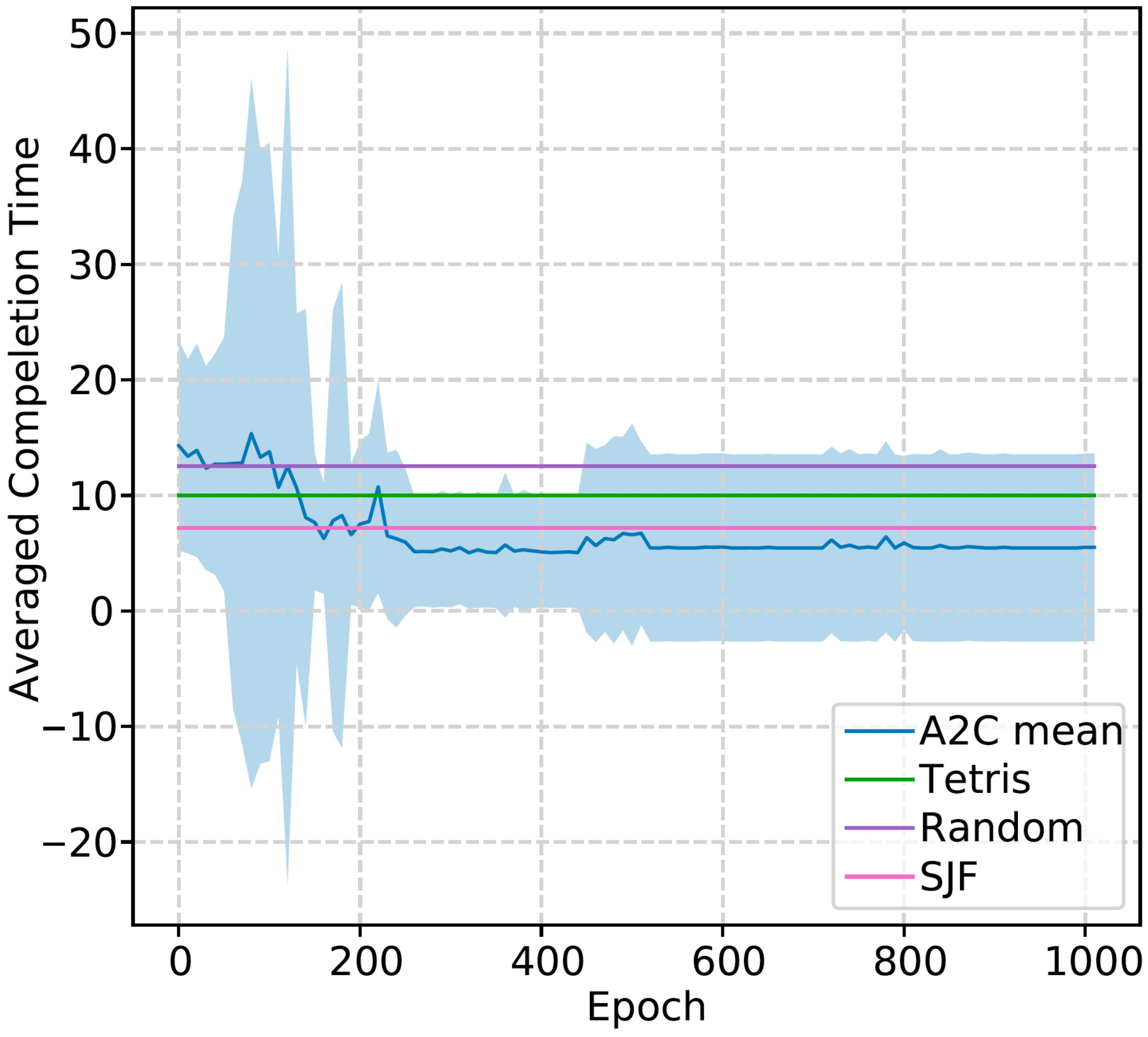}
	    \label{fig:learning_completion_time0.9}
	}
	\subfigure[Average waiting time.]{
	    \includegraphics[width=0.4\linewidth]{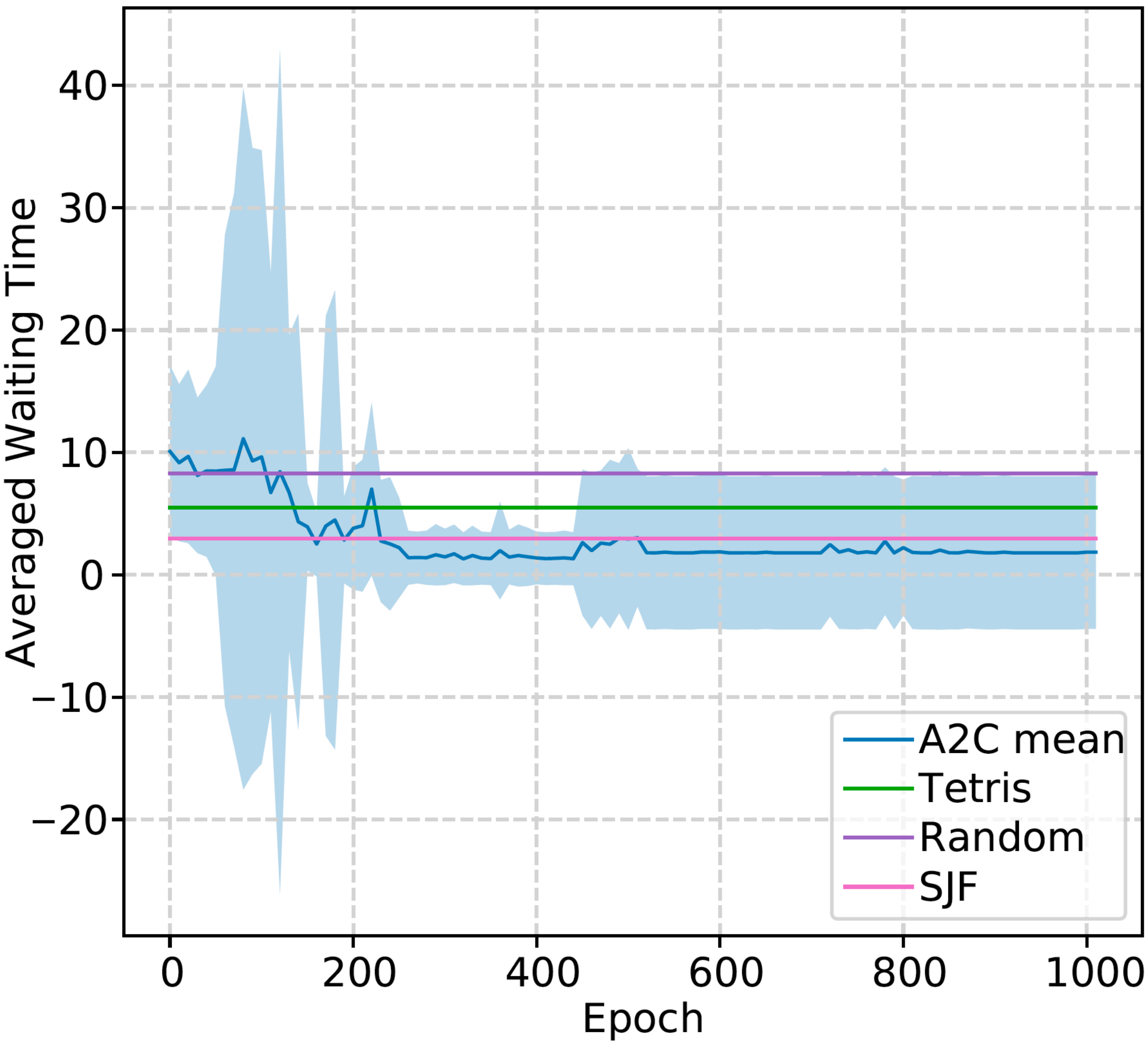}
	    \label{fig:learning_waiting_time_0.9}
	}
	\caption{A2C performance with a \textit{job arrival rate=0.7} } 
    \label{fig:my_label2}
\end{figure*}


\begin{figure*}[t]
    \centering
	\subfigure[Discounted reward.]{
	    \includegraphics[width=0.4\linewidth]{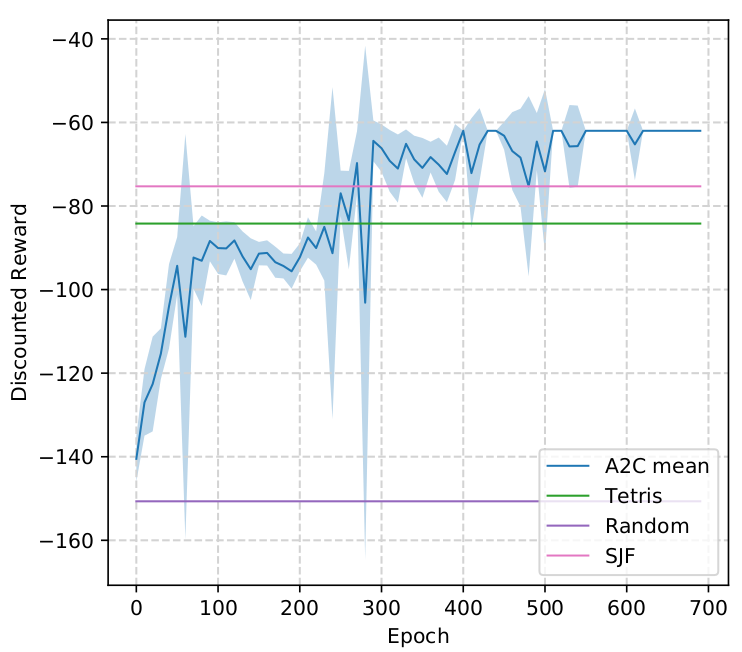}
	    \label{fig:log_reward}
	}
	\subfigure[Slowdown.]{
	    \includegraphics[width=0.4\linewidth]{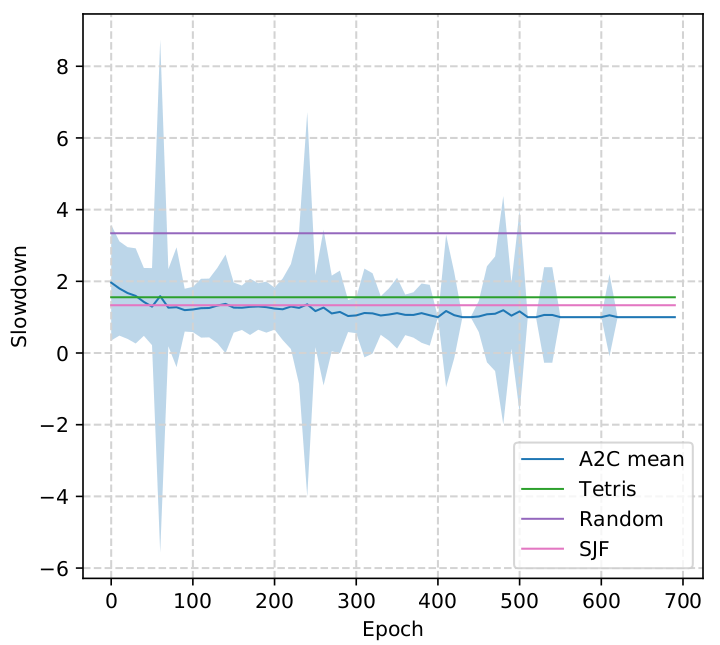}
	    \label{fig:log_slown_down}
	}
	\subfigure[Average completion time.]{
	    \includegraphics[width=0.4\linewidth]{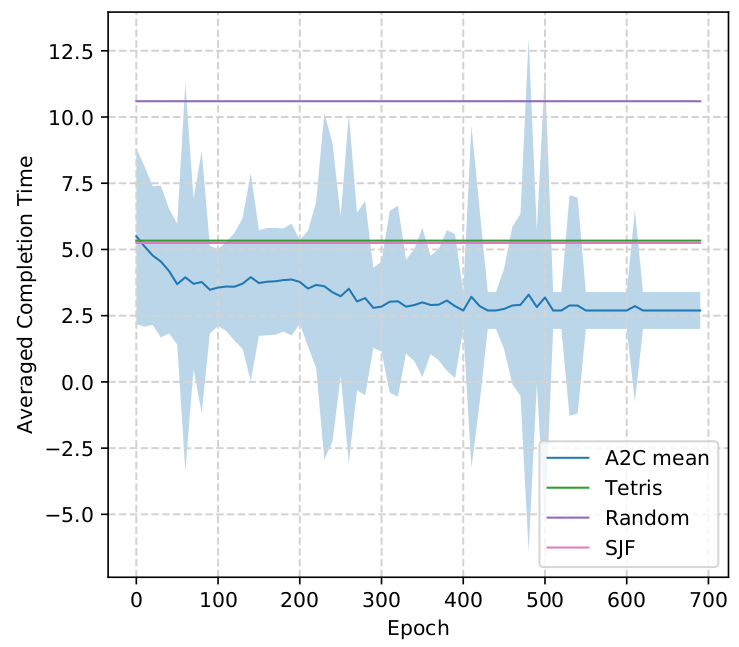}
	    \label{fig:log_completion_time}
	}
	\subfigure[Average waiting time.]{
	    \includegraphics[width=0.4\linewidth]{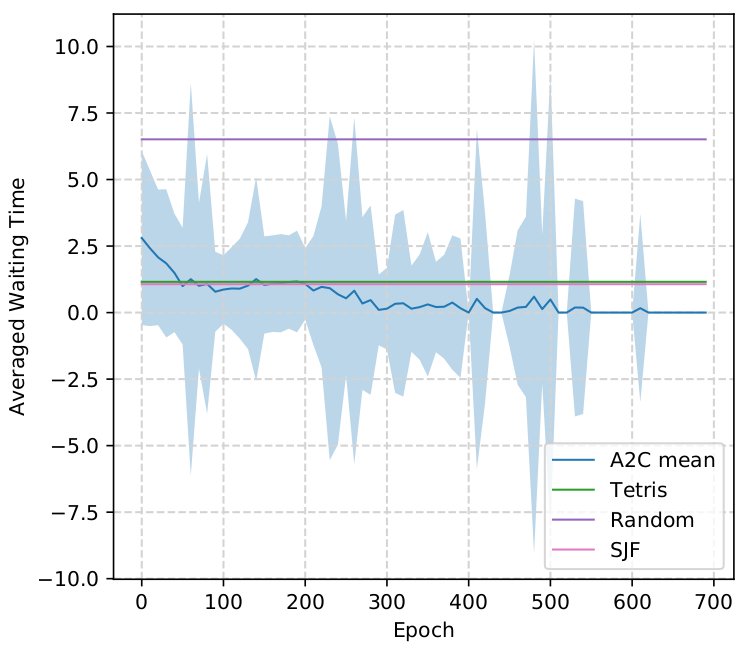}
	    \label{fig:log_waiting_time}
	}
	\caption{A2C performance with real world log data } 
    \label{fig:log_label}
\end{figure*}





\subsection{A2cScheduler with CNN}
We simulated the data center cluster containing $N$ nodes with two resources: CPU and Memory. 
We trained the A2cScheduler with different neural networks including a fully connected layer and Convolutional Neural Networks (CNN). In order to design the best performance neural networks, we explore different CNN architectures and compare whether it converges and how is the converge speed with different settings. 
As shown in table~\ref{tab:addlabel}, fully connected layer (FC layer) with a flatten layer in front did not converge. This is because the state of the environment is a matrix with location information while some location information lost in the flatten layer when the state is processed. 
To keep the location information, we utilize CNN layers (16 3*3-filters CNN layer  and 32 3*3-filters CNN layer) and they show better results. Then, we explored CNN with max-pooling and CNN with flattening layer behind. Results show both of them could converge but CNN with max-pooling gets poorer results. This is due to some of the state information also get lost when it passes max-pooling layer. According to the experiment results, we decide to choose the CNN with a flattening layer behind architecture as it converges fast and gives the best performance. 

\begin{table}[!htbp]
\small
  \centering
  \caption{{Performances of different network architectures.}}
    \begin{tabular}{p{10.555em}p{7.541em}p{7.55em}p{7.71em}}
    \toprule\toprule
    \multicolumn{1}{l}{Architecture} & \multicolumn{1}{l}{Converge} &  \multicolumn{1}{l}{Converging}  & \multicolumn{1}{l}{Converging} \\
          &       & Speed  &  Epochs \\
    \midrule
    FC layer & No    & N.A.  & N.A. \\
    \midrule
    Conv3-16 & Yes   & Fast  & 500 \\
    \midrule
    Conv3-32 & Yes   & Slow  & 1100 \\
    \midrule
    Conv3-16 + pooling & Yes & Fast& 700 \\
    \midrule
    Conv3-32 + pooling & Yes & Fast & 900 \\
    \bottomrule\bottomrule
    \end{tabular}%
  \label{tab:addlabel}%
\end{table}%

\begin{table}[!hbtp]
\small
  \centering
  \caption{{Results of Job Traces.}}
  
    \begin{tabular}{p{6.555em}p{6.541em}p{6.55em}p{6.71em}p{6.565em}}
    \toprule\toprule
    Type  & \multicolumn{1}{l}{Random} & \multicolumn{1}{l}{Tetris} & \multicolumn{1}{l}{SJF} & \multicolumn{1}{l}{A2cScheduler} \\
    \midrule
    Slowdown & 3.52$\pm$0.00 & 1.82$\pm$0.00 & 1.61$\pm$0.00 & \textbf{1.01$\pm$0.02} \\
    \midrule
    $CT^*$    & 10.2$\pm$0.00 & 5.55$\pm$0.00 & 5.51$\pm$0.00 & \textbf{2.58$\pm$0.01} \\
    \midrule
    $WT^*$ & 6.32$\pm$0.00 & 1.25$\pm$0.00 & 1.21$\pm$0.00 & \textbf{0.01$\pm$0.02} \\
    \bottomrule\bottomrule
    \end{tabular}%
  \label{tab:converged_hpcc}
\end{table}%

\subsection{Baselines}







The performance of the proposed method is compared with some of the mainstream baselines such as
Shortest Job First (SJF), Tetris~\cite{grandl2015multi}, and random policy. SJF sorts jobs according to their execution time
and schedules jobs with the shortest execution time first; Tetris schedules job by a combined score of preferences
for the short jobs and resource packing; random policy schedules jobs randomly. All of these baselines work in a greedy
way that allocates as many jobs as allowed by the resources, and share the same resource constraints and take 
the same input as the proposed model. 

\subsection{Performance Comparison}
\paragraph{\textbf{Performance on Synthetic Dataset}}
In our experiment, the A2cScheduler utilized an A2C reinforcement learning method. 
It is worth to mention that the model includes the option to have multiple episodes in order to allow us to measure the average performance achieved and the capacity to learn for each scheduling policy. Algorithm \ref{A2C} presents the calculation and update of parameters per episode. 
Figure~\ref{fig:my_label2} shows experimental results with synthetic job distribution as input.

Figure~\ref{fig:learning_reward_0.9} and Figure~\ref{fig:learning_slown_down0.9} present the rewards and averaged slowdown when the new job rate is 0.7. Cumulative rewards and averaged slowdown converge around 500 episodes. A2cScheduler has lower averaged slowdown than random, Tetris and SJF after 500 episodes.
Figure~\ref{fig:learning_completion_time0.9} and Figure~\ref{fig:learning_waiting_time_0.9} show the average completion time and average waiting time of	the A2cScheduler	algorithm versus baselines. As we can see, the performance of A2cScheduler is the best comparing to all the baselines. 

Table~\ref{tab:converged_98},~\ref{tab:converged_76} present the steady state simulation results at different job rates. We can see the A2cScheduler algorithm gets the best or close to the best performance regrading slowdown, average completion time and average waiting time at different job rates ranging from 0.6 to 0.9.

\paragraph{\textbf{Performance on Real-world Dataset}}
We ran experiments with a piece of job trace from an academic data center. The results were shown in figure \ref{fig:log_label}. The job traces were preprocessed before they are trained with the A2cScheduler. There was some fluctuation at the first 500 episodes in \ref{fig:log_reward}, then it started to converge. Figure \ref{fig:log_slown_down} shows the average slowdown was better than all the baselines and close to optimal value 1, which means the average waiting time was almost 0 as shown in figure \ref{fig:log_waiting_time}. This happens because there were only 60 jobs in this case study and jobs runtime are small. This was an case where almost no job was waiting for the allocation when it was optimally scheduled. A2cScheduler also gains the shortest completion time among different methods from figure~\ref{fig:log_completion_time}. Table~\ref{tab:converged_hpcc} shows the steady state results from a real-world job distribution running on an academic cluster. A2cScheduler gets optimal scheduling results since there is near 0 average waiting time for this jobs distribution. Again, this experimental results proves A2cScheduler effectively finds the proper scheduling policies by itself given adequate training, both on simulation dataset and real-world dataset. 
There were no rules predefined for the scheduler in advance, instead, there was only a reward defined with the system optimization target included. This proven our defined reward function was effective in helping the scheduler to learn the optimal strategy automatically after adequate training.


\section{Conclusion}
Job scheduling with resource constraints is a long-standing but critically important problem for computer systems. In this paper, we proposed an A2C deep reinforcement learning algorithm to address the customized job scheduling problem in data centers
We defined a reward function related to averaged job waiting time which leads A2cScheduler to find scheduling policy by itself. 
Without the need for any predefined rules, this scheduler is able to automatically learn strategies directly from experience and thus improve scheduling policies. 
Our experiments on both simulated data and real job traces for a data center show that our proposed method performs better than widely used SJF and Tetris for multi-resource cluster scheduling algorithms, offering a real alternative to current conventional approaches. 
The experimental results reported in this paper are based on two-resource (CPU/Memory) restrictions, but this approach can also be easily adapted for more complex multi-resource restriction scheduling scenarios. 

\section{Acknowledgement}
We are thankful to the anonymous reviewers for their valuable feedback. This research is supported in part by the National Science Foundation under grant CCF-1718336 and CNS-1817094.

\bibliographystyle{splncs04}
\bibliography{reference}

\begin{thebibliography}{10}
\providecommand{\url}[1]{\texttt{#1}}
\providecommand{\urlprefix}{URL }
\providecommand{\doi}[1]{https://doi.org/#1}

\bibitem{al2007model}
Al-Tamimi, A., et~al.: Model-free q-learning designs for linear discrete-time
  zero-sum games with application to h-infinity control. Automatica
  \textbf{43}(3),  473--481 (2007)

\bibitem{domeniconi2019cush}
Domeniconi, G., Lee, E.K., Morari, A.: Cush: Cognitive scheduler for
  heterogeneous high performance computing system  (2019)

\bibitem{garg2011sla}
Garg, S.K., Gopalaiyengar, S.K., Buyya, R.: Sla-based resource provisioning for
  heterogeneous workloads in a virtualized cloud datacenter. In: Proc. ICA3PP.
  pp. 371--384 (2011)

\bibitem{grandl2015multi}
Grandl, R., et~al.: Multi-resource packing for cluster schedulers. Computer
  Communication Review  \textbf{44}(4),  455--466 (2015)

\bibitem{grondman2012survey}
Grondman, I., et~al.: A survey of actor-critic reinforcement learning: Standard
  and natural policy gradients. IEEE Transactions on Systems, Man, and
  Cybernetics  \textbf{42}(6),  1291--1307 (2012)

\bibitem{hovestadt2003scheduling}
Hovestadt, M., Kao, O., Keller, A., Streit, A.: Scheduling in hpc resource
  management systems: Queuing vs. planning. In: Workshop on JSSPP. pp. 1--20.
  Springer (2003)

\bibitem{konda2000actor}
Konda, V.R., et~al.: Actor-critic algorithms. In: Proc. NIPS. pp. 1008--1014
  (2000)

\bibitem{liu2018reinforcement}
Liu, Z., Zhang, H., Rao, B., Wang, L.: A reinforcement learning based resource
  management approach for time-critical workloads in distributed computing
  environment. In: Proc. Big Data. pp. 252--261. IEEE (2018)

\bibitem{mao2018learning}
Mao, H., et~al.: Learning scheduling algorithms for data processing clusters.
  arXiv preprint arXiv:1810.01963  (2018)

\bibitem{Mao:2016:RMD:3005745.3005750}
Mao, H., Alizadeh, M., Menache, I., Kandula, S.: Resource management with deep
  reinforcement learning. In: HotNets '16. pp. 50--56. ACM, New York.
  \doi{10.1145/3005745.3005750}

\bibitem{mnih2016asynchronous}
Mnih, V., et~al.: Asynchronous methods for deep reinforcement learning. In:
  Proc. ICML. pp. 1928--1937 (2016)

\bibitem{sprague2003multiple}
Sprague, N., Ballard, D.: Multiple-goal reinforcement learning with modular
  sarsa (0)  (2003)

\bibitem{srinivasan2018actor}
Srinivasan, S., et~al.: Actor-critic policy optimization in partially
  observable multiagent environments. In: Proc. NIPS. pp. 3422--3435 (2018)

\bibitem{sutton1998introduction}
Sutton, R.S., Barto, A.G., et~al.: Introduction to reinforcement learning,
  vol.~135. MIT press Cambridge (1998)

\bibitem{van2012scheduling}
Van~Craeynest, K., et~al.: Scheduling heterogeneous multi-cores through
  performance impact estimation (pie). In: Computer Architecture News. vol.~40,
  pp. 213--224

\bibitem{watkins1992q}
Watkins, C.J., Dayan, P.: Q-learning. Machine learning  \textbf{8}(3-4),
  279--292 (1992)

\bibitem{yangcoordinating}
Yang, Z., Nguyen, L., Jin, F.: Coordinating disaster emergency response with
  heuristic reinforcement learning

\bibitem{zhou2013exploring}
Zhou, X., Chen, H., Wang, K., Lang, M., Raicu, I.: Exploring distributed
  resource allocation techniques in the slurm job management system. Technical
  Report  (2013)

\end{thebibliography}

\end{document}